\documentclass[fp,twocolumn]{jpsj3}
\usepackage{txfonts}
\usepackage{color}
\usepackage{graphicx}
\usepackage{epstopdf}
\usepackage{amsmath}
\usepackage{amsfonts}
\usepackage{amssymb}
\usepackage[version=3]{mhchem}

\title{
Elastic Soft Mode and Electric Quadrupole Response in Excitonic Insulator Candidate (Ta$_{0.952}$V$_{0.048}$)$_2$NiSe$_5$: Contribution of Electron-Phonon Interaction
}

\author{
Ryosuke Kurihara$^1$\thanks{r.kurihara@issp.u-tokyo.ac.jp}, 
Yusuke Hirose$^2$,
Sumika Sano$^3$,
Keisuke Mitsumoto$^4$,
Atsushi Miyake$^1$, \\
Masashi Tokunaga$^1$,
and
Rikio Settai$^2$
}

\inst{
$^1$The Institute for Solid State Physics, The University of Tokyo, Kashiwa, Chiba 277-8581, Japan \\
$^2$Department of Physics, Niigata University, Niigata 950-2181, Japan \\
$^3$Graduate School of Science and Technology, Niigata University, Niigata 950-2181, Japan \\
$^4$Liberal Arts and Sciences, Toyama Prefectural University, Imizu, Toyama 939-0398, Japan
} 

\abst{
We performed ultrasonic measurements on (Ta$_{0.952}$V$_{0.048}$)$_2$NiSe$_5$ to investigate the contribution of acoustic phonons at the zone center to structural and possible excitonic transitions.
The transverse elastic constant $C_{55}$ with irreducible representation $B_\mathrm{2g}$ exhibits softening over 90\% above a transition temperature of $T_\mathrm{s} = 280$ K, indicating the soft mode of corresponding TA phonons.
Our analysis based on the quadrupole response indicates that the electron-phonon interaction contributes to the excitonic transition in addition to the hybridization between the conduction and valence bands if the excitonic transition is realized.
}

\begin{document}
\maketitle

\section{Introduction}

Structural phase transition (SPT) and the instability of related phonons are accompanied by several phase transitions and orderings of electronic degrees of freedom, e.g., ferroelectric transitions and charge, orbital, and multipole orderings
\cite{Rehwald_AdvPhys22, Goto_Luthi_review, Tokura_Science288, Luthi_Textbook}.
In addition, recent studies in iron-based compounds have shown the relation between the structural instability caused by the orbital or multipole fluctuations and superconductivity
\cite{Ono_SSCommun152, Kontani_PRB84}.
Thus, the investigation of the origin of SPT and phonon instability can help discover exotic phenomena.

The excitonic insulator transition (EIT) is one of the exotic phase transitions.
EIT is described by particle pairs comprising electrons in the conduction band and holes in the valence band exhibiting the Bose-Einstein-type condensation in semiconductor and the BCS-type condensation in semimetal
\cite{Cloizeaux_JPCS26, Kohn_PRL19, Halperin_RMP40, Bronold_PRB74}.
Owing to condensation, the instability of conduction and valence bands occurs.
In addition to the instability of the electronic structure, SPT accompanied by the EIT has been proposed in Ta$_2$NiSe$_5$ and 1$T$-TiSe$_2$
\cite{Wakisaka_PRL103, Kaneko_PRB87, Pillo_PRB61, Kogar_Science358, Mazza_PRL124}.
Superconductivity caused by the suppression of SPT in these materials
\cite{Kusmartseva_PRL103, Morosan_NPhys2}
has seemed to attract more attention to EIT.

In EIT, theoretical predictions have revealed that several characteristic properties of superconductivity appear because the Hamiltonian and the order parameter of EIT, $\langle v^\dagger c \rangle$ ($\langle c^\dagger v \rangle$),  are similar to that of superconducting transition.
Here, $v^\dagger$ ($v$) is the creation (annihilation) operator of particles on valence bands and $c$ ($c^\dagger$) is the annihilation (creation) operator of particles on conduction bands.
One of the important phenomena expected for the EIT is the coherence peak. 
In the BCS-type superconductors, the coherence peak appears in the temperature dependence of the spin-lattice relaxation rate $1/T_1$ just below the superconducting transition temperature.
On the other hand, the appearance of the coherence peak in EIT has been expected in the ultrasonic attenuation coefficient \cite{Maki_JLTP5, Amritkar_SSC26}.
Therefore, the existence of the coherence peak in ultrasonic attenuation coefficients strongly indicates the EIT transition.
In particular, a theoretical study considering the electron-phonon interaction for the EIT accompanying SPT has proposed the importance of ultrasonic measurements in the EIT candidate compound Ta$_2$NiSe$_5$
\cite{Sugimoto_PRB93}.

\begin{figure}[b]
\begin{center}
\includegraphics[clip, width=0.5\textwidth, bb=0 0 600 330]{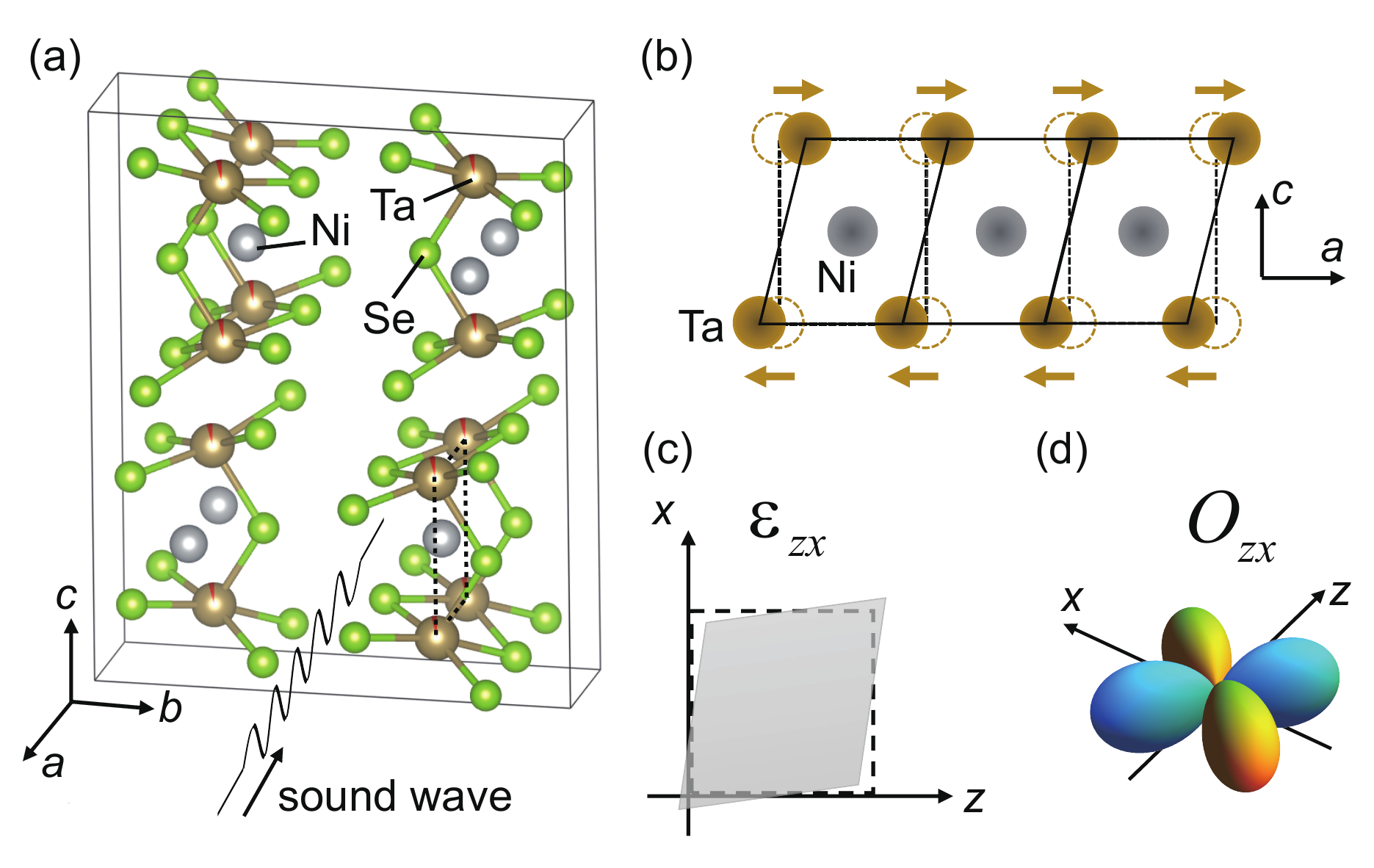}
\end{center}
\caption{
(Color online)
(a) Crystal structure of Ta$_2$NiSe$_5$ produced with VESTA
\cite{VESTA}.
Sound waves for the elastic constant $C_{55}$ are schematically shown.
(b) Schematic view of the lattice distortion due to the SPT.
The dashed rectangular indicates the non-distorted structure using 4 Ta-ions.
This is also exhibited in (a).
The parallelogram shows the deformed structure.
The horizontal arrows indicate the displacement direction of Ta ions.
(c) Strain $\varepsilon_{zx}$ and (d) electric quadrupole $O_{zx}$ with the irrep $B_\mathrm{2g}$. 
}
\label{Fig1}
\end{figure}

The crystal structure of Ta$_2$NiSe$_5$ belonging to the $Cmcm$ ($D_{2h}^{17}$) orthorhombic space group
\cite{Sunshine_IChem24}
is shown in Fig. \ref{Fig1}(a).
Below the SPT temperature $T_\mathrm{s} = 328$ K, crystal symmetry is lowered to the monoclinic $C2/c$ $(C_{2h}^6)$ that satisfies group-subgroup relation 
\cite{InternationalTables, DiSalvo_JLCM116, Nakano_PRB98}.
Owing to this orthorhombic-to-monoclinic SPT, rectangular lattices on $ac$-plane formed by 4 Ta-ions deform to parallelograms, as depicted in Fig. \ref{Fig1}(b).
This lattice deformation is described by the strain $\varepsilon_{zx}$ shown in Fig. \ref{Fig1}(c).

From the electronic structure point of view, Ta-$5d$ and Ni-$3d$ electrons seem to play a key role in the SPT and EIT.
In Ta$_2$NiSe$_5$, several studies have shown that the electronic system has a direct gap with the gap minimum around the $\Gamma$ point of the Brillouin zone center
\cite{Kaneko_PRB87, Wakisaka_PRL103}.
Near the Fermi level, the valence band ($v$-band) and the conduction band ($c$-band) consist of the Ta-$5d_{xy}$ and Ni-$3d_{zx+yz}$ orbitals.
Theoretical proposals for EIT are based on these bands around the $\Gamma$ point
\cite{Kaneko_PRB87, Mazza_PRL124, Yamada_JPSJ85}.
ARPES studies have observed that these bands exhibit spontaneous flattening below $T_\mathrm{s}$ indicating the EIT
\cite{Wakisaka_PRL103, Seki_PRB90}.
These results indicate that the order parameter of the SPT is carried by $d$ electrons.

As the contribution of the electron-phonon interaction to EIT has been considered
\cite{Kaneko_PRB87, Sugimoto_PRB93},
the determination of the soft mode of SPT is also significantly investigated.
Based on the inelastic x-ray scattering measurements, the softening of acoustic phonons with the irreducible representation (irrep) $B_\mathrm{2g}$ in $\Gamma$ - $Z$ - $\Gamma$ direction has been observed above $T_\mathrm{s}$
\cite{Nakano_PRB98}.
The instability of zone-center optical phonon mode with the irrep $B_\mathrm{2g}$ has also been proposed as the soft mode by Raman spectroscopy and the theoretical calculations of phonon dispersion
\cite{Kim_PRR2, Subedi_PRM4}. 
On the other hand, a theoretical study has proposed elastic softening in the transverse elastic constant $C_{55}$, which corresponds to TA phonons at the zone center and uniform structural deformation (see Figs. \ref{Fig1}(b) and (c))
\cite{Sugimoto_PRB93}.
Several studies by Raman spectroscopy have also predicted the softening of $C_{55}$ and related sound velocity
\cite{Kim_Ncommun12, Ye_PRB104}.

From these studies, elastic measurements based on ultrasonic methods seem to be important to investigate SPT and EIT in Ta$_2$NiSe$_5$.
The analysis of the elastic softening provides the energy scale of electron-phonon interaction for acoustic phonons.
Furthermore, owing to the typical frequency range of $f \sim 10^7$ - $10^8$ Hz and wavelength range of $\lambda \sim 10^{-6}$ - $10^{-4}$ m, ultrasonic measurement is a good probe to observe acoustic phonons with low-energy and small wave vectors.
However, ultrasonic measurements have not been available because the single crystal size of Ta$_2$NiSe$_5$ was not sufficient to perform ultrasonic pulse-echo methods.
Recently, some of the authors have developed the single crystal growth process for isovalent V-substituted (Ta$_{1-x}$V$_x$)$_2$NiSe$_5$
\cite{Sano_JPSConfProc30}.
They succeeded in growing large single crystals capable of ultrasonic measurements.
This substituted system shows the same SPT as that of Ta$_2$NiSe$_5$ except for the SPT temperature of $T_\mathrm{s} = 280$ K.
In addition, the profile of the temperature dependence of resistivity rate in (Ta$_{1-x}$V$_x$)$_2$NiSe$_5$ is similar to that of Ta$_2$NiSe$_5$.
Thus, we deduce that the physical properties of SPT, possible EIT, and the topology of the band structure in (Ta$_{1-x}$V$_x$)$_2$NiSe$_5$ are the same as Ta$_2$NiSe$_5$.
Here, we investigate elastic softening owing to the structural instability around $T_\mathrm{s}$ and coherence peak in the ultrasonic attenuation coefficient caused by EIT.

This paper is organized as follows.
In Sect. \ref{Experiments}, the experimental procedures of sample preparation and ultrasonic measurements are described.
In Sect. \ref{Results}, we present the results of the ultrasonic experiments that demonstrate the soft mode elastic constant $C_{55}$ and the softening of related TA phonons at zone center caused by the SPT in (Ta$_{0.952}$V$_{0.048}$)$_2$NiSe$_5$.
In Sect. \ref{Discussions}, we discuss the energy scale of the electron-phonon interaction in terms of the electric quadrupole response describing elastic softening of $C_{55}$.
We also consider the electric quadrupole degrees of freedom originate from the spin-singlet pairs formed by the EIT.

\section{Experiments}
\label{Experiments}

Single crystals of (Ta$_{1-x}$V$_x$)$_2$NiSe$_5$ with the nominal concentration of $x = 0.20$ were grown using the chemical transport method that includes iodine as a transport agent
\cite{Sano_JPSConfProc30}.
The actual amount of V ions, $x = 0.048$, was estimated using wavelength-dispersive spectroscopy with Electron Probe Micro Analyzer
\cite{Sano_JPSConfProc30}.
Laue x-ray backscattering was used to align crystallographic orientations.
We carefully polished the sample with alumina abrasive powder with \#1000 (Maruto Instrument Co.) and liquid paraffin (FUJIFILM Wako Pure Chemical Co.) to obtain (100), ($\bar{1}$00), (001), and (00$\bar{1}$) surfaces for ultrasonic measurement.
(010) and (0$\bar{1}$0) surfaces were cleavage planes.
As SPT corresponds to ferroelasticity, external strains easily change the elastic property of crystals as proposed in the iron pnictide compound
\cite{Zhang_PRB94}.
One of the samples was deformed owing to the wrong amount of force generated by a pair of tweezers.
This mechanical weakness of (Ta$_{1-x}$V$_x$)$_2$NiSe$_5$ is reflected in the small absolute value of transverse elastic constants $C_{55}$ and $C_{66}$ with the order of $10^{9}$ J/m$^3$.
If the entire sample is covered with adhesive, unexpected external strain can be introduced into the sample owing to these differences in the thermal expansion coefficients.
Therefore, we fixed the sample on the probe using weak-type double-sided tape (NICHIBAN Co) to avoid introducing bulk strain to the sample. 
The sample size dimension 2.05 mm ($a$ axis) $\times$ 0.587 mm ($b$ axis) $\times$ 1.45 mm ($c$ axis) was sufficient to perform the ultrasonic measurements. 

The ultrasonic pulse-echo method with a numerical vector-type phase-detection technique was used to measure the ultrasonic velocity $v_{ij}$ and the ultrasonic attenuation coefficient $\alpha_{ij}$
\cite{Fujita_JPSJ80}.
The absolute value of $v_{ij}$ was determined to read the time interval between the cross talk and 1st echo signals (see Fig. \ref{Echo}$\cdot1$ in Appendix \ref{EchoSignals}).
The elastic constant $C = \rho v_{ij}^2$ was determined from $v_{ij}$ and the mass density $\rho = $ 7.652 g/cm$^3$ for the V concentration $x = 0.048$, which is estimated using the lattice constants $a = 3.482 \mathrm{\AA}$, $b = 12.80 \mathrm{\AA}$, and $c = 15.61 \mathrm{\AA}$
\cite{Hirose_Private}.
The resolution of the elastic constants was estimated to be about $10^{-4}$.
Longitudinal and transverse ultrasonic waves were obtained using 36$^\circ$ Y- and X-cut LiNbO$_3$ piezoelectric transducers with the 200- and 100-micrometer thickness (YAMAJU CO).
Typical fundamental frequencies of 36$^\circ$ Y- and X-cut plates were 33 MHz and 17 MHz, respectively.
X-cut plates with the 200-micrometer thickness were also used in the $C_{55}$ measurements around the SPT temperature by the fundamental frequency of 9 MHz.
The wave length of ultrasonic waves, $\lambda = f/v_{ij}$, in our measurements was estimated to be $5 \times 10^{-5}$ - $1 \times 10^{-4}$ m$^{-1}$ by the ultrasonic frequency $f$ and sound velocity $v_{ij}$.
The dicing size of 36$^\circ$ Y- and X-cut LiNbO$_3$ was 1.3 mm $\times$ 1.3 mm and 1.5 mm $\times$ 1.3 mm, respectively.
Here, the polarization direction of X-cut LiNbO$_3$ was the short side of the plates.
The transducers were pasted onto the sample with glue (TOAGOSEI Co., Aron Alpha jelly type).
The thickness of glue was estimated to be less than $6$ $\mu$m.
Because this value was negligibly smaller than the sample size, we ignored the thickness to calculate sound velocity.

$^4$He cryostat was used to measure the temperature dependence of elastic constants and ultrasonic attenuation coefficients below RT.
The Quantum Design Physical Property Measurement System with hand-made ultrasonic measurement systems was used to measure the elastic constants $C_{55}$ and $C_{66}$ above RT up to 390 K. 

\section{Results}
\label{Results}

\begin{figure*}[t]
\begin{center}
\includegraphics[clip, width=0.8\textwidth, bb=0 0 600 550] {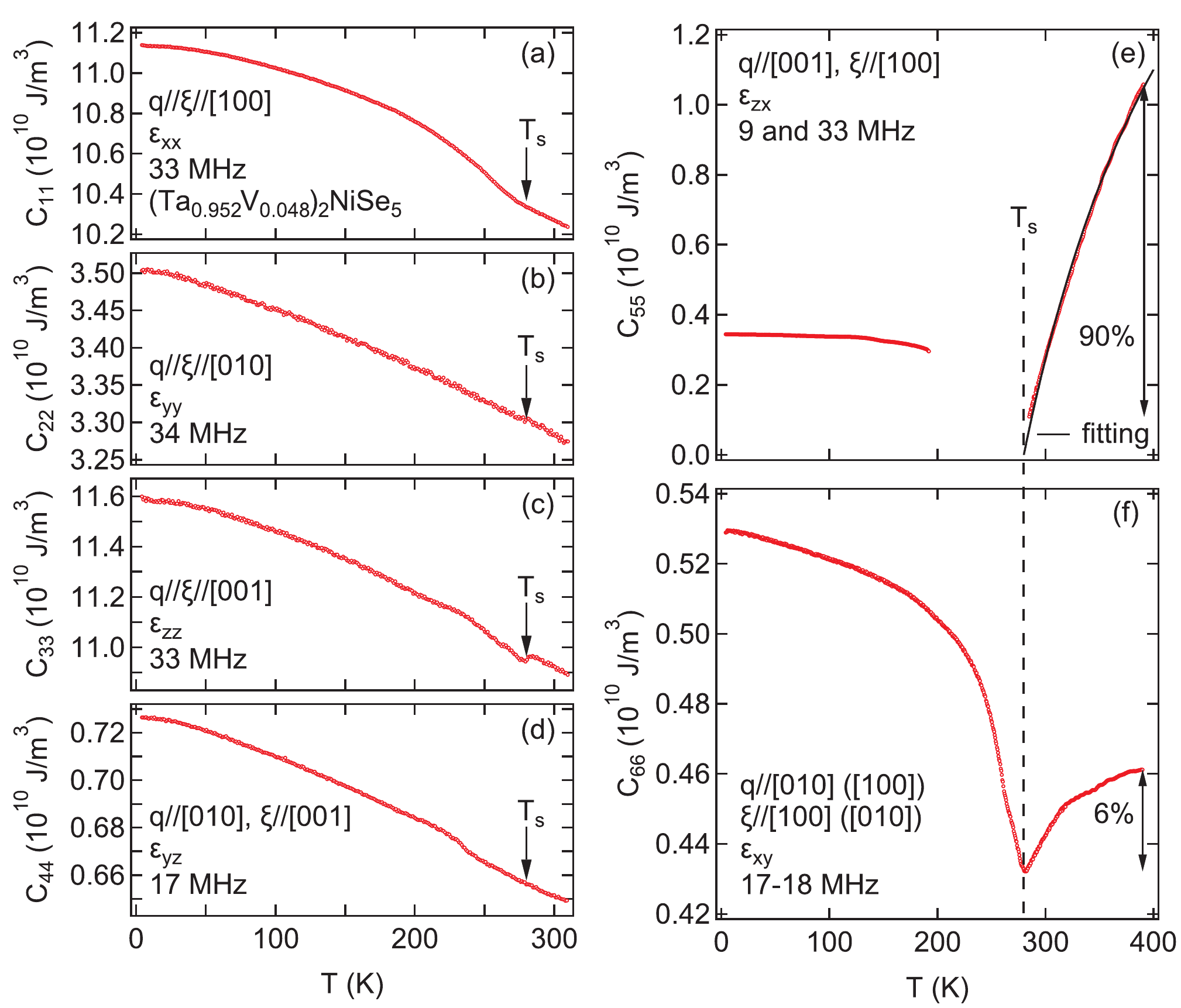}
\end{center}
\caption{
(Color online)
Temperature dependence of (a) the longitudinal elastic constants $C_{11}$, (b) $C_{22}$, and (c) $C_{33}$, and  the transverse elastic constants (d) $C_{44}$, (e) $C_{55}$, and (f) $C_{66}$ in (Ta$_{0.952}$V$_{0.048}$)$_2$NiSe$_5$.
The ultrasonic propagation direction $\boldsymbol{q}$ and polarization direction $\boldsymbol{\xi}$, induced strains $\varepsilon_{ij}$, and used frequency of each measurement are also listed.
The solid line in panel (e) indicates the fit of $C_{55}$ in the framework of the phenomenological form of Eq. (\ref{Curie}).
The down arrows and the dashed line in each panel indicate $T_\mathrm{s}$.
The double-headed arrows show the amount of softening from 390 K down to 285 K in $C_{55}$ and down to $T_\mathrm{s}$ in $C_{66}$. 
}
\label{C_Tdep}
\end{figure*}
\begin{table*}[ht]
\caption{
Symmetry strains, electric multipoles, elastic constants, and basis functions corresponding to the irreducible representations (irreps) of the $D_{2h}$
\cite{Inui_group}.
}
\label{table1}
\begin{tabular}{cccccc}
\hline
\textrm{Irrep}
& \textrm{Symmetry strain}
	& \textrm{Electric multipole}
		& \textrm{Elastic constant}
			&\textrm{Basis function}
\\
\hline
$A_\mathrm{1g}$
	& $\varepsilon_\mathrm{B} = \varepsilon_{xx} + \varepsilon_{yy} + \varepsilon_{zz}$
		& $O_0$
			& $ C_\mathrm{B} = \left( C_{11} + C_{22} + C_{33} + 2C_{12} + 2C_{13} + 2 C_{23} \right) /9$
				&$x^2$, $y^2$, $z^2$
\\
 
	& $\varepsilon_u = (2\varepsilon_{zz} - \varepsilon_{xx} - \varepsilon_{yy})/\sqrt{3}$
		& $O_{u(=3z^2-r^2)}$
			& $C_u = \left( C_{11} + C_{22} + 4C_{33} + 2C_{12} - 4C_{13} - 4C_{23} \right)/12$
				&
\\

	& $\varepsilon_v = \varepsilon_{xx} - \varepsilon_{yy}$
		& $O_{v(=x^2-y^2)}$
			& $C_v = \left( C_{11} + C_{22} -2C_{12} \right)/4$
				&
\\
$B_\mathrm{1g}$
	& $\varepsilon_{xy}$
		& $O_{xy}$
			& $C_{66}$
				&$xy$
\\
$B_\mathrm{2g}$
	& $\varepsilon_{zx}$
		& $O_{zx}$
			& $C_{55}$
				&$zx$
\\
$B_\mathrm{3g}$
	& $\varepsilon_{yz}$
		& $O_{yz}$
			& $C_{44}$
				& $yz$
\\
\hline
\end{tabular}
\end{table*}

To identify the soft mode of the SPT in (Ta$_{0.952}$V$_{0.048}$)$_2$NiSe$_5$, we investigated six elastic constants, $C_{11}$, $C_{22}$, $C_{33}$, $C_{44}$, $C_{55}$, and $C_{66}$, which can be measured in the orthorhombic structure.
The irrep of reduced elastic constants and corresponding strains in $D_{2h}$ are listed in Table \ref{table1}.
Figure \ref{C_Tdep} shows the temperature dependence of elastic constants $C_{ij}$.
The longitudinal elastic constants, $C_{11}$, $C_{22}$, and $C_{33}$ exhibit weak anomalies at $T_\mathrm{s} = 280$ K.
The transverse elastic constant $C_{44}$ does not show elastic softening.
In contrast, the transverse elastic constant $C_{55}$ with the irrep $B_\mathrm{2g}$ (see Table \ref{table1}) demonstrates over $90\%$ softening from 390 K down to 285 K.
Owing to this huge softening, we were not able to obtain the reproducible data of $C_{55}$ below 285 K down to $\sim 190$ K.
This can be attributed to the scattering of ultrasonic waves caused by structural domains.
The difficulty of elastic measurements around SPT has been reported in (CH$_3$NH$_3$)$_2$MnCl$_4$, (CH$_3$NH$_3$)$_2$CdCl$_4$, and CeCu$_6$
\cite{Goto_PRB22, Goto_JMMM63}.
Below 190 K, $C_{55}$ shows monotonic hardening with decreasing temperature.
As the transverse elastic constant $C_{66}$ exhibits softening above $T_\mathrm{s}$, the SPT induces the strain $\varepsilon_{xy}$ with the irrep $B_\mathrm{1g}$.
However, it is not dominant because the amount of softening from 390 K down to $T_\mathrm{s}$, estimated to be 6\%, is much smaller than that of $C_{55}$.
These experimental results confirm that $C_{55}$ is the soft mode elastic constant of the SPT in (Ta$_{0.952}$V$_{0.048}$)$_2$NiSe$_5$.
Therefore, we conclude that the TA phonons with the irrep $B_\mathrm{2g}$ at the zone center freeze at $T_\mathrm{s}$, in agreement with the inelastic x-ray and Raman scattering measurements in Ta$_2$NiSe$_5$
\cite{Nakano_PRB98, Kim_PRR2}. 

To observe the coherence peak caused by the EIT, we tried to measure the temperature dependence of the ultrasonic attenuation coefficients in (Ta$_{0.952}$V$_{0.048}$)$_2$NiSe$_5$.
However, we cannot confirm the coherence peak around the SPT in the ultrasonic attenuation coefficient $\alpha_{55}$ of the transverse ultrasonic wave for the soft mode elastic constant $C_{55}$ because of the scattering of the ultrasonic waves.
We also tried ultrasonic attenuation measurements in other ultrasonic waves.
The ultrasonic attenuation coefficient $\alpha_{11}$ of the longitudinal elastic constant $C_{11}$ with the ultrasonic frequency of 104 MHz and 175 MHz showed a broad peak around 278 K, which was slightly lower than $T_\mathrm{s}$.
The amount of increase of $\alpha_{11}$ from 280 K down to 278 K was about 100 dB/m, however, the signal-noise ratio was comparable to 100 dB/m.
Therefore, we could not conclude whether the coherence peak just below $T_\mathrm{s}$ exists or not.
We also could not observe the peak structure in other ultrasonic attenuation coefficients.
Even though the coherence peak owing to the EIT has theoretically been predicted in $\alpha_{55}$ in Ta$_2$NiSe$_5$, we suppose that it is hard to observe the peak because of the scattering of ultrasonic waves around $T_\mathrm{s}$.
Theoretical predictions of the coherence peak in the longitudinal ultrasonic waves are desired in Ta$_2$NiSe$_5$.

\section{Discussions}
\label{Discussions}

For further understanding of the SPT in (Ta$_{1-x}$V$_{x}$)$_2$NiSe$_5$, we discuss the electric quadrupole contributions to the elastic softening of $C_{55}$ in terms of the phenomenological theory.
The SPT from the $Cmcm$ space group to that of the subgroup $C2/c$ allows the active representation of the irrep $B_\mathrm{2g}$.
Therefore, the order parameter of SPT is described by the $zx$-type basis function
\cite{Landau_Lifshitz},
as listed in Table \ref{table1}.
This is also shown by the point group analysis (see Appendix \ref{PointGroup}).
Thus, we can describe the SPT using the electric quadrupole $O_{zx}$ with the irrep $B_\mathrm{2g}$ shown in Fig. \ref{Fig1}(d) as an order parameter.
The phenomenological free energy is described below
\cite{Luthi_Textbook}:
\begin{align}
\label{Free energy}
F = \frac{1}{2} \alpha_0 &(T - \mathit{\Theta}) O_{zx}^2 + \frac{1}{4} \beta_0 O_{zx}^4 \nonumber
\\
		& - g_{zx} O_{zx}\varepsilon_{zx}  + \frac{1}{2}C_{55}^0 \varepsilon_{zx}^2
.
\end{align}  
Here, $\alpha_0$ and $\beta_0$ are the positive coefficients, $\mathit{\Theta}$ is the quadrupole ordering temperature without the strain $\varepsilon_{zx}$, $g_{zx}$ is the quadrupole-strain coupling constant, and $C_{55}^0$ is the elastic constant without the contribution of quadrupole to free energy.
The 1st and 2nd terms in the right-hand side of Eq. (\ref{Free energy}) describe the quadrupole ordering of $O_{zx}$. 
These terms originate from the quadrupole-quadrupole interaction Hamiltonian with the mean-field approximation
\cite{Nishimori_Textbook}.
The 3rd term is known as the quadrupole-strain interaction that describes the energy gain owing to the strain $\varepsilon_{zx}$.
From a microscopic point of view, the quadrupole-strain interaction coincides with the decomposed electron-phonon interaction.
The 4th term is the elastic energy caused by the crystal deformation.
The equilibrium condition of the free energy of Eq. (\ref{Free energy}),
$\partial F/\partial O_{zx} = 0$,
and the 2nd derivative of the free energy with respect to the strain $\varepsilon_{zx}$ provide the temperature dependence of $C_{55}$.
\begin{align}
\label{Curie}
C_{55} = C_{55}^0
 \left(
1 - \frac{ \mathit{\Delta} }{ T - \mathit{\Theta} }
\right)
= 
C_{55}^0 \left(
\frac{T - T_\mathrm{s}^0 }{ T - \mathit{\Theta} }
\right)
.
\end{align}
Here, $\mathit{\Delta}$ is known as the Jahn-Teller energy, including the quadrupole-strain coupling constant, and $T_\mathrm{s}^0 = \mathit{\Delta} + \mathit{\Theta}$ is the theoretical SPT temperature.
Comparing the microscopic form of quadrupole susceptibility
\cite{Kurihara_JPSJ86},
we obtain $\mathit{\Delta}$ = $N g_{zx}^2/C_{55}^0$ where $N$ is the number of particles per unit volume possessing quadrupole.
In addition, $\mathit{\Theta}$ can be understood as the Weiss temperature that corresponds to the strength of the quadrupole-quadrupole interaction.
The elastic formula of Eq. (\ref{Curie}) reproduces the softening of $C_{55}$ in (Ta$_{0.952}$V$_{0.048}$)$_2$NiSe$_5$ as shown in Fig. \ref{C_Tdep}(e), except for around $T_\mathrm{s}$ because it is hard to measure an elastic constant around SPT owing to the huge attenuation of ultrasound.
This difficulty can be attributed to the critical-slowing-down like iron pnictide compounds
\cite{Kurihara_JPSJ86}. 

\begin{table}[t]
\caption{
Characteristic parameters and values in (Ta$_{0.952}$V$_{0.048}$)$_2$NiSe$_5$
}
\label{table_parameters}
\begin{tabular}{cc}
\hline
Parameters
	& Values
\\
\hline
$\mathit{\Theta}$
	& $97$ K
\\
$\mathit{\Delta}$
	& $183$ K
\\
$T_\mathrm{s}^0 = \mathit{\Delta} + \mathit{\Theta}$
	& $280$ K
\\
$C_{55}^0$
	& $2.78 \times 10^{10}$ J/m$^3$
\\
$N$ (Ni)
	& $5.75\times 10^{27}$ m$^{-3}$
\\
$N$ (Ta)
	& $1.15\times 10^{28}$ m$^{-3}$
\\
$g_{zx}$ for Ni-ions
	& $2530$ K
\\
$g_{zx}$ for Ta-ions
	& $5660$ K
\\
\hline
\end{tabular}
\end{table}

To reproduce the softening of $C_{55}$ with Eq. (\ref{Curie}), we obtain characteristic parameters describing the SPT of (Ta$_{0.952}$V$_{0.048}$)$_2$NiSe$_5$ as shown in Table \ref{table_parameters}.
These results provide several important pieces of information about the SPT and EIT.
One is the contribution of quadrupole-strain interaction, namely the electron-phonon interaction, to the SPT.
The positive value of $\mathit{\Theta}$ indicates the ferro-type quadrupole-quadrupole interaction, suggesting that the electronic system can spontaneously exhibit ferro-type quadrupole ordering at $T = \mathit{\Theta} = 97$ K without the quadrupole-strain interaction.
This result indicates the importance of electronic correlations for the SPT.
On the other hand, our analysis demonstrates the finite contribution of quadrupole-strain interaction to the SPT in terms of $\mathit{\Delta} = 183$ K that raises the transition temperature from $\mathit{\Theta} = 97$ K to $T_\mathrm{s}^0 = 280$ K.
Therefore, we conclude that the quadrupole-strain interaction, namely the electron-phonon interaction for acoustic phonons, can be necessary to drive the SPT in (Ta$_{0.952}$V$_{0.048}$)$_2$NiSe$_5$.
Furthermore, if EIT accompanying SPT is realized in (Ta$_{0.952}$V$_{0.048}$)$_2$NiSe$_5$, the EIT temperature cannot be $\mathit{\Theta} = 97 $ K, which can be related to the proposed putative EIT temperature
\cite{Kim_Ncommun12},
but $T_\mathrm{s}^0 = 280$ K as also proposed by Raman spectroscopy
\cite{Volkov_njpQM6}.
Another ingredient to the SPT is the contribution of $d$ electrons.
Using $\mathit{\Delta} = 183$ K, $C_{55}^0 = 2.78 \times 10^{10}$ J/m$^3$, and the number of Ni (Ta)-ions per unit volume, $N =  5.750 \times 10^{27}$ $(1.150 \times 10^{28})$ m$^{-3}$, we estimate the quadrupole-strain coupling constant $g_{zx}$ to be  $2530$ $(5660) $ K.
In the microscopic point of view, the spatially expanded wave functions provide a large coupling constant because $g_{zx}$ is proportional to $\langle \psi_i | O_{zx}  | \psi _j \rangle$ where $\psi_i$ is the wave function describing electronic systems
\cite{Kurihara_JPSJ86}.
In fact, $g \sim 1000$ - $3000$ K is comparable to that of several materials exhibiting orbital or quadrupole orderings in $d$-electron systems
\cite{Hazama_PRB62, Hazama_PRB69, Goto_JPSJ80, Yoshizawa_JPSJ81, Kurihara_JPSJ86} but more than 10 times larger than that of $4f$ electron systems
\cite{Nakamura_JPSJ63}.
Therefore, these facts indicate that the Ni-$3d$ and/or Ta-$5d$ electrons contribute to the quadrupole degree of freedom of $O_{zx}$ and the SPT in (Ta$_{0.952}$V$_{0.048}$)$_2$NiSe$_5$.
Owing to this strong quadrupole-strain interaction, the huge elastic softening of $C_{55}$ is induced by the SPT despite the small crystal deformation that is described by the angle change between $z$- and $x$-axis from $ 90^\circ$ to $\sim 90.5^\circ$
\cite{DiSalvo_JLCM116}.

Furthermore, we discuss the relation between our results in (Ta$_{0.952}$V$_{0.048}$)$_2$NiSe$_5$ and previous theoretical studies for Ta$_2$NiSe$_5$. 
The elastic soft mode $C_{55}$ observed in the present study agrees with the theoretical prediction
\cite{Sugimoto_PRB93}.
Thus, our ultrasonic experiments do not contradict the SPT accompanied by the EIT as theoretically proposed.
The quadrupole-strain interaction, $-g_{zx}O_{zx}\varepsilon_{zx}$, can be attributed to the electron-phonon interaction for the structural deformation describing the hybridization between the conduction and valence bands mediated by phonons
\cite{Kaneko_PRB87, Sugimoto_PRB93}.
Therefore, the presence of the quadrupole-strain interaction with the coupling constant of 2530 - 5660 K indicates that not only the inter-particle interaction between the valence and conduction bands but also the electron-phonon interaction for zone-center TA phonons contribute to EIT in (Ta$_{0.952}$V$_{0.048}$)$_2$NiSe$_5$ if the EIT is realized.
The bare electron-phonon coupling constant can be obtained from the quadrupole-strain coupling constant to calculate the expectation value of $O_{zx}$, $\langle \psi_i | O_{zx}  | \psi _j \rangle$, for the wave functions $\psi_i$ describing SPT and EIT.
We expect that these results stimulate more theoretical predictions about the mechanism of EIT and SPT in Ta$_2$NiSe$_5$.

Our ultrasonic measurements and analysis indicate the contribution of the electric quadrupole $O_{zx}$ to the SPT in (Ta$_{0.952}$V$_{0.048}$)$_2$NiSe$_5$.
To investigate the possibility of EIT-induced quadrupole ordering, we discuss the origin of $O_{zx}$ based on the group theory and band structure of Ta$_2$NiSe$_5$.
The order parameter of EIT is proportional to $\langle c^\dagger v  \rangle$ ($\langle v^\dagger c \rangle$) for the spin-singlet pairing
\cite{Kaneko_PRB87}.
On the other hand, the expectation value of the second quantization form of electric multipole, $\langle O_\Gamma \rangle$, between the particles on the $v$- and $c$-bands is written as 
$\langle \psi_c| O_\Gamma | \psi_v \rangle \langle c^\dagger v  \rangle$
($\langle \psi_v | O_\Gamma | \psi_c \rangle \langle v^\dagger c  \rangle$)
\cite{Kurihara_JPSJ86}. 
Here, $\psi_v$ ($\psi_c$) is the wave function describing particles on the $v$- ($c$-) band and $\Gamma$ denotes the suffix of electric quadrupoles. 
If the SPT accompanied by the EIT is realized in Ta$_2$NiSe$_5$, both
$\langle \psi_c | O_\Gamma | \psi_v \rangle$ ($\langle \psi_v | O_\Gamma | \psi_c \rangle$)
and  $\langle c^\dagger v  \rangle$ ($\langle v^\dagger c  \rangle$) show non-zero value.
The wave function of particles on the $v$- and $c$-bands consist of Ta-$5d_{xy}$ orbital with the irrep $B_\mathrm{1g}$ and Ni-$3d_{zx+yz}$ orbital with $B_\mathrm{2g} \oplus B_\mathrm{3g}$ in the $D_{2h}$ group (see Table \ref{table1}).
Based on the group-theoretical analysis, we show that the multipole degrees of freedom,
$\langle \psi_c | O_\Gamma | \psi_v \rangle$ ($\langle \psi_v | O_\Gamma | \psi_c \rangle$),
carried by the spin-singlet pairs are obtained by reducing 
$B_\mathrm{1g} \otimes (B_\mathrm{2g} \oplus B_\mathrm{3g})$ in Ta$_2$NiSe$_5$.
As shown in Table \ref{table_Product}, the result of this reduction turns out to be $B_\mathrm{2g}$ (= $B_\mathrm{1g} \otimes B_\mathrm{3g}$) and $B_\mathrm{3g}$  (= $B_\mathrm{1g} \otimes B_\mathrm{2g}$).
This group-theoretical analysis demonstrates that the spin-singlet pairs describing the EIT also provides the electric quadrupole $O_{zx}$ with the irrep $B_\mathrm{2g}$ and the non-zero value of $\langle O_{zx} \rangle$ can be obtained in the EIT phase.
Since $\langle O_{zx} \rangle$ is proportional to $\langle c_{yz}^\dagger v_{xy}  \rangle$ ($\langle v_{xy}^\dagger c_{yz} \rangle$), the order parameter of the EIT can partly be written by $\langle O_{zx} \rangle$.
Therefore, the transition temperature of the EIT, which can correspond to the quadrupole ordering temperature of $O_{zx}$ at $\mathit{\Theta}$, increases to $T_\mathrm{s}^0$ owing to the electron-phonon interaction then SPT is induced.
We deduce that this result based on the group theory can be an example of the electric quadrupole ordering induced by EIT theoretically predicted
\cite{Kaneko_PRB94}.
Mulitpole orderings can be considered in excitonic-ordered materials with multi-orbital bands. 

We need to implement  high-temperature measurements to reveal the starting temperature of the elastic softening and obtain reliable parameters providing static properties of the quadrupole.
Moreover, $C_{55}$ and $\alpha_{55}$ measurements below $T_\mathrm{s}$, i.e., down to $\sim 190$ K, is required.
Further challenges detecting coherence peak around $T_\mathrm{s}$ are very important to understand EIT.
The ultrasonic measurements in a non-substituted Ta$_2$NiSe$_5$ are needed to escape from the disorder effect in the substituted systems.
Since the huge ultrasonic attenuation around the SPT is observed in the elastic soft mode $C_{55}$, it may be hard to measure the coherence peak in the soft mode.
Further theoretical prediction of the coherence peak in the different ultrasonic attenuation coefficients is highly desired.

\begin{table}[t]
\caption{
Product table for gerade irreps of $D_{2h}$
\cite{Inui_group}
}
\label{table_Product}
\begin{tabular}{c|cccc}
\hline
 
	&$A_\mathrm{1g}$
		&$B_\mathrm{1g}$
			&$B_\mathrm{2g}$
				&$B_\mathrm{3g}$
\\
\hline
$A_\mathrm{1g}$
	&$A_\mathrm{1g}$
		&$B_\mathrm{1g}$
			&$B_\mathrm{2g}$
				&$B_\mathrm{3g}$
\\
$B_\mathrm{1g}$ 
	&$B_\mathrm{1g}$
		&$A_\mathrm{1g}$
			&$B_\mathrm{3g}$
				&$B_\mathrm{2g}$
\\
$B_\mathrm{2g}$ 
	&$B_\mathrm{2g}$
		&$B_\mathrm{3g}$
			&$A_\mathrm{1g}$
				&$B_\mathrm{1g}$
\\
$B_\mathrm{3g}$ 
	&$B_\mathrm{3g}$
		&$B_\mathrm{2g}$
			&$B_\mathrm{1g}$
				&$A_\mathrm{1g}$
\\
\hline
\end{tabular}
\end{table}

\section{Conclusion}

In conclusion, we investigated elastic soft mode accompanied by the SPT of (Ta$_{0.952}$V$_{0.048}$)$_2$NiSe$_5$ based on ultrasonic measurements.
The transverse elastic constant $C_{55}$ for the response of strain $\varepsilon_{zx}$ with the irrep $B_\mathrm{2g}$ exhibits huge softening over 90\% on approaching $T_\mathrm{s}$, while other elastic constants do not exhibit such huge softening.
Therefore, we conclude that $C_{55}$ is the soft mode of the SPT and corresponding TA phonons at the zone center freeze at $T_\mathrm{s}$.
We were able to reproduce the softening of $C_{55}$ in the framework of phenomenological analysis, describing electric quadrupole response.
The Jahn-Teller energy, $\mathit{\Delta} = 183 $ K, and estimated coupling constant, $g_{zx} = 2530$ - $5660$ K, indicate the strong coupling of the strain $\varepsilon_{zx}$ with the quadrupole $O_{zx}$ originating from $d$ electrons.
In other words, corresponding TA phonons with irrep $B_\mathrm{2g}$ strongly interact with $d$ electrons.
These facts indicate that the electron-phonon interaction in addition to the hybridization between the $c$- and $v$-bands plays a key role in the EIT in (Ta$_{0.952}$V$_{0.048}$)$_2$NiSe$_5$ if the SPT accompanied by EIT is realized.
Furthermore, our experimental results can confirm the theoretical prediction of electric quadrupole ordering induced by EIT that can be a new exotic mechanism describing SPT.

\begin{acknowledgment}

The authors thank Yoshiaki \=Ono, Takuya Sekikawa, and Koudai Sugimoto for valuable discussions.
This work was supported by Grants-in-Aid for young scientists (KAKENHI 	JP20K14414).
This work was also partly supported by Grants-in-Aid for young scientists (KAKENHI JP20K14404) and scientific research (KAKENHI JP20K03854, JP19K03713, and JP19H00648).

\end{acknowledgment}


\appendix

\section{Determination of sound velocity}
\label{EchoSignals}

\begin{figure}[t]
\label{Echo}
\begin{center}
\includegraphics[clip, width=0.5\textwidth, bb=0 0 380 650] {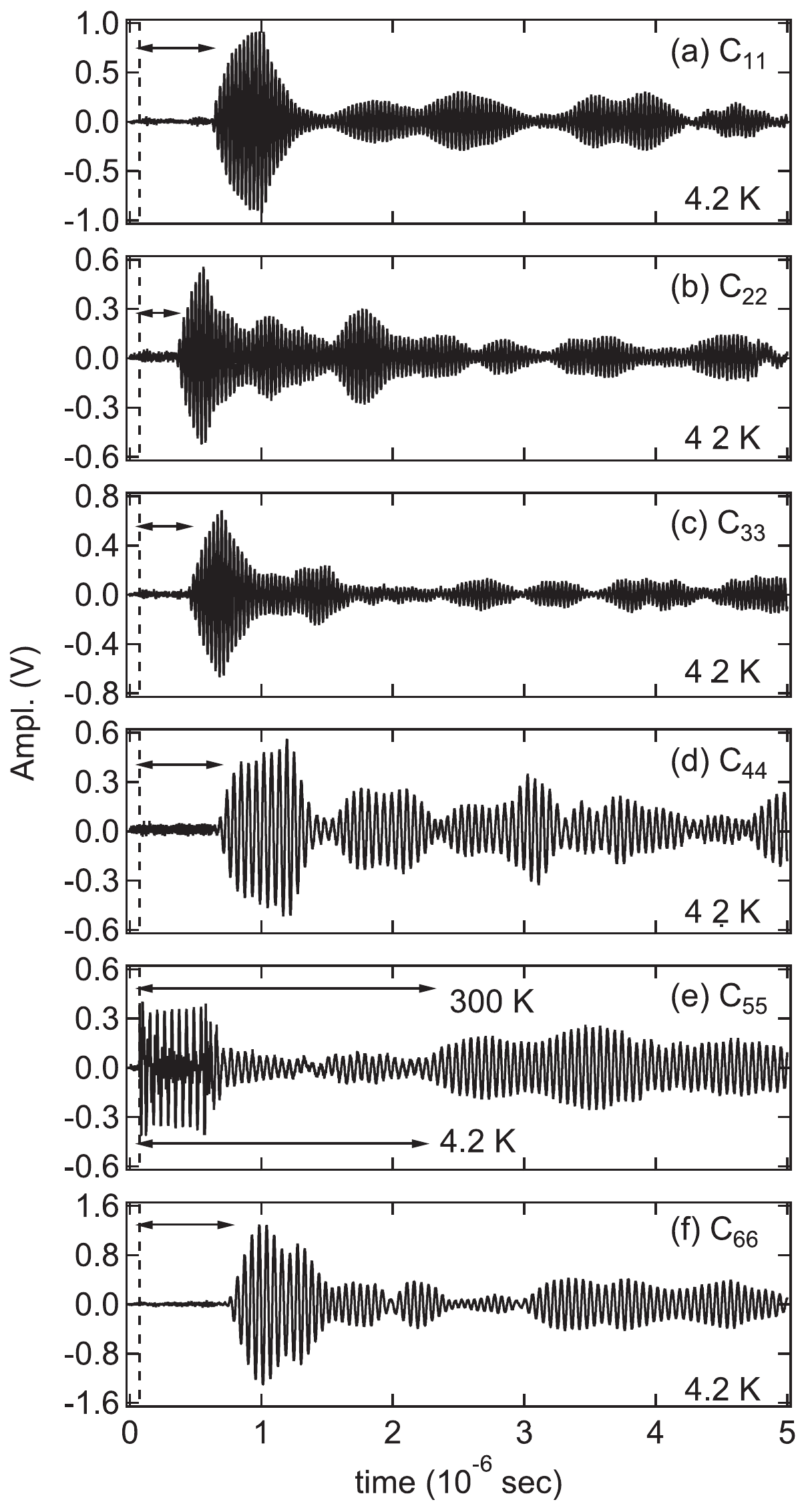}
\end{center}
\caption{
Time dependence of the amplitude of ultrasonic echo signals for the elastic constants in (Ta$_{0.952}$V$_{0.048}$)$_2$NiSe$_5$.
The vertical broken lines in each panel show the starting point of the cross-talk signal.
A typical cross-talk signal is exhibited in panel (e).
The horizontal arrows indicate the time interval between the cross-talk and 1st echo signals.
}
\end{figure}
Figure \ref{Echo}$\cdot1$ shows the time dependence of the ultrasonic echo signals in (Ta$_{0.952}$V$_{0.048}$)$_2$NiSe$_5$.
In the pulse-echo methods, the pulse-modulated electronic signals are converted to ultrasonic waves by the piezoelectric plate.
After the propagation of ultrasonic waves in the sample, ultrasonic waves are again converted to electronic signals by another piezoelectric plate.
On the other hand, there are to be cross-talk signals directly propagating to the piezoelectric plate on the opposite side.
Therefore, the propagation time of ultrasonic waves can be determined by the time interval between the 1st echo and cross-talk signals.
In Fig. \ref{Echo}$\cdot1$, we show the time interval between the 1st echo and cross-talk signals.
Dividing the propagation length of ultrasonic waves by the time interval, we calculated the sound velocity $v_{ij}$. 
Since the time interval is determined to read echo signals, the error in $v$ is considered to be about 10\%.
Because the adhesive width between the sample and LiNbO$_3$ piezoelectric plates was estimated to be much shorter than the sample length, we ignored the adhesive width to calculate the absolute value of $v_{ij}$.

\section{Point group analysis}
\label{PointGroup}

As elastic constants and ultrasonic attenuation coefficients reflect the global properties of the crystals, the point group analysis can help understand the SPT and phenomenological free energy in Ta$_2$NiSe$_5$.
Tables \ref{table_C2v} and \ref{table_C2} show the characters and basis functions
\cite{Inui_group}
of point group $C_{2v}$ and $C_2$ at Ni-ion sites
\cite{Sunshine_IChem24}
, respectively.
Here, $E$ is the notation for the identity operation, $C_2$ is the $\pi$-rotation operation around $y$ axis, $\sigma_x$ is the mirror operation for $yz$ plane, and $\sigma_z$ is the mirror operation for $xy$ plane.
The irrep $A_2$ of the point group $C_{2v}$ belongs to the full symmetry $A$ of the point group $C_2$.
Therefore, we confirm that the irrep $A_2$ of $C_{2v}$ is the active representation that describes symmetry lowering from $C_{2v}$ to $C_2$.
This fact shows that the phenomenological free energy for the SPT is written by the $zx$-type order parameter.
Furthermore, this group-theoretical analysis can be consistent with the possible species of ferroelastic crystals
\cite{Aizu_PRB2}.

\begin{table}[htbp]
\caption{
Characters and basis functions of point group $C_{2v}$.
}
\label{table_C2v}
\begin{tabular}{cccccc}
\hline
\textrm{irrep}
	& $E$
		& $\sigma_x$
			& $\sigma_z$
				& $C_2$
					& \textrm{basis function}
\\
\hline
$A_1$
	& $1$
		& $1$
			& $1$
				& $1$
					& $y, x^2, y^2, z^2$
\\
$A_2$
	& $1$
		& $-1$
			& $-1$
				& $1$
					& $zx$
\\
$B_1$
	& $1$
		& $1$
			& $-1$
				& $-1$
					& $z, yz$
\\
$B_2$
	& $1$
		& $-1$
			& $1$
				&$-1$
					& $x, xy$
\\
\hline
\end{tabular}
\caption{
Characters and basis functions of point group $C_{2}$.
}
\label{table_C2}
\begin{tabular}{cccc}
\hline
\textrm{irrep}
	& $E$
		& $C_2$
			& \textrm{basis function}
\\
\hline
$A$
	& $1$
		& $1$
			& $y, x^2, y^2, z^2, zx$
\\
$B$
	& $1$
		& $-1$
			& $x, z, xy, yz$
\\
\hline
\end{tabular}
\end{table}


\end{document}